\documentclass[aps,twocolumn,superscriptaddress]{revtex4}
\usepackage{epsfig}
\usepackage{times}
\usepackage{color}

\begin{document}

\title{Beyond pairwise strategy updating in the prisoner's dilemma game}

\author{Xiaofeng Wang}
\affiliation{Center for Complex Systems, Xidian University, Xi'an 710071, China}

\author{Matja{\v z} Perc}
\email{matjaz.perc@uni-mb.si}
\affiliation{Faculty of Natural Sciences and Mathematics, University of Maribor, Koro{\v s}ka cesta 160, SI-2000 Maribor, Slovenia}

\author{Yongkui Liu}
\affiliation{School of Automation Science and Electrical Engineering, Beihang University, Beijing 100191, China}
\affiliation{School of Electronic and Control Engineering, Chang'an University, Xi'an 710054, China}
\affiliation{Center for Road Traffic Intelligent Detection and Equipment Engineering, Chang'an University, Xi'an 710054, China}

\author{Xiaojie Chen}
\affiliation{Evolution and Ecology Program, International Institute for Applied Systems Analysis (IIASA), Schlossplatz 1, A-2361 Laxenburg, Austria}

\author{Long Wang}
\affiliation{Center for Systems and Control, State Key Laboratory for Turbulence and Complex Systems, College of Engineering, Peking University, Beijing 100871, China}

\begin{abstract}
In spatial games players typically alter their strategy by imitating the most successful or one randomly selected neighbor. Since a single neighbor is taken as reference, the information stemming from other neighbors is neglected, which begets the consideration of alternative, possibly more realistic approaches. Here we show that strategy changes inspired not only by the performance of individual neighbors but rather by entire neighborhoods introduce a qualitatively different evolutionary dynamics that is able to support the stable existence of very small cooperative clusters. This leads to phase diagrams that differ significantly from those obtained by means of pairwise strategy updating. In particular, the survivability of cooperators is possible even by high temptations to defect and over a much wider uncertainty range. We support the simulation results by means of pair approximations and analysis of spatial patterns, which jointly highlight the importance of local information for the resolution of social dilemmas.
\end{abstract}

\maketitle

Cooperative behavior is extremely important, both in the animal world as well as across human societies \cite{axelrod_84, bowles_11, hrdy_11, nowak_11}. Yet it is also an evolutionary puzzle, as it is costly but has no immediate individual benefits, except in rare exceptions, for example when cooperation is agreed upon as a risk-sharing mechanism. How cooperation evolved amongst selfish and unrelated individuals is therefore still ardently investigated, as evidenced by recent reviews \cite{doebeli_el05, nowak_s06, szabo_pr07, schuster_jbp08, roca_plr09, perc_bs10}.

Evolutionary game theory \cite{hofbauer_98, nowak_06, sigmund_10} provides an apt theoretical framework to address the subtleties of the evolution of cooperation. One of the most popular games that is representative for situations constituting a social dilemma is the prisoner's dilemma game \cite{axelrod_84}. It can be summarized succinctly. Two individuals have to decide simultaneously whether they wish to cooperate or not. Cooperator pays a cost $c$ towards the mutual benefit
$b$ where $b>c>0$, while defector contributes nothing. This yields the temptation to defect $T=b$, reward for mutual cooperation $R=b-c$, punishment for mutual defection $P=0$, and the sucker's payoff $S=-c$, which for the prisoner's dilemma game thus satisfy $T>R>P>S$ and $2R>T+S$. Evidently, for an individual it is best to defect regardless of what the opponent does. As rational players are aware of this, they both defect, in turn obtaining $P$ rather than $R$, hence the social dilemma \cite{glance_sa94}.

Several mechanisms that facilitate the evolution of cooperation are known. Nowak summarizes five
rules \cite{nowak_s06}, which are kin selection \cite{hamilton_wd_jtb64b}, direct reciprocity \cite{trivers_qrb71}, indirect reciprocity \cite{nowak_n98}, group selection \cite{wilson_ds_pnas75}, and network reciprocity \cite{nowak_n92b}. Networks in particular, have received substantial attention in the recent past \cite{szabo_pr07}. While scale-free networks appear to provide the best environment for the evolution of cooperation \cite{santos_prl05, santos_pnas06, santos_prsb06, gomez-gardenes_prl07, poncela_njp07, szolnoki_pa08, poncela_epl09, perc_njp09}, small-world \cite{santos_pre05, ren_pre07, fu_epjb07, perc_njp06c, chen_xj_pre08} and
hierarchical networks \cite{vukov_pre05, gomez-gardenes_jtb08, lee_s_prl11} also received ample attention. Largely motivated by the discovery that complex networks facilitate the evolution of cooperation, heterogeneity in general has emerged as an important property that may help keep defectors in the minority \cite{santos_n08, perc_pre08, perc_njp11, santos_jtb12}. Coevolutionary games \cite{perc_bs10}, where the structure of the network was subject to evolution just as the strategies of players have been studied thoroughly too \cite{zimmermann_pre04, zimmermann_pre05, pacheco_jtb06, pacheco_prl06, santos_ploscb06, fu_pre08b, fu_pre09, chen_xj_pre09b, wu_t_epl09, szolnoki_epl08, poncela_ploso08, poncela_njp09, szolnoki_epl09, szolnoki_njp09, zhang_cy_pone11}, with the prevailing conclusion being that this may give rise to robust cooperative states and lead to socially preferable interaction networks in a spontaneous manner. Quite remarkably, this has recently been confirmed empirically \cite{rand_pnas11}, although very extensive experiments also indicate that the human behavior may suppress network reciprocity \cite{gracia-lazaro_srep12, gracia-lazaro_pnas12}.

In fact, how human decision-making affects the evolution of cooperation is of particular relevance for the present work. Szab{\'o} et al. \cite{szabo_pre10} have recently considered a special type of strategy updating. Instead of players exclusively caring only about their own
payoffs when updating their strategies, they investigated what happens when a pair of randomly chosen neighboring players tries to maximize their collective income by simultaneously updating their two strategies. It was reported that the proposed strategy update rule produces the antiferromagnetic ordering structure of cooperators and defectors on the square lattice at sufficiently low noise intensities, and that this favors the evolution of cooperation more than the traditional pairwise imitation updating. Human decision-making dynamics has also been investigated experimentally, whereby we are particularly interested in the so called ``social influence'' effect reported by Lorenz et al. \cite{lorenz_pnas11}. As stated in their paper, social influence among group members plays an important role in individual decision-making.

One may then ask how this affects the evolution of cooperation? To address this question, we propose an adaptive strategy-adoption rule in which the social influence is taken into account. In particular, as a proxy for the social influence we assume that the decisions the players make are affected by all their neighbors, not just a single randomly selected or the most successful neighbor. Players can collect information from their neighbors, and moreover, their decision-making is more likely to be affected by the circle of ``close friends'' rather than the whole social environment. We introduce this so-called local influence to the strategy updating simply so that, before a potential update, each player considers the average performance of its own strategy and that of the other strategy, if present, within its neighborhood. As we will show in what follows, this introduces a qualitatively different evolutionary dynamics that is able to support the stable existence of very small cooperative clusters, which in turn supports the survivability of cooperative behavior even under very unfavorable conditions. Besides simulation results \cite{huberman_pnas93}, we will also present results obtained with pair approximation methods, which are, along with the game theoretical model, accurately described in the Methods section.

\section*{Results}

\begin{figure}
\centering
\includegraphics[width=8.5cm]{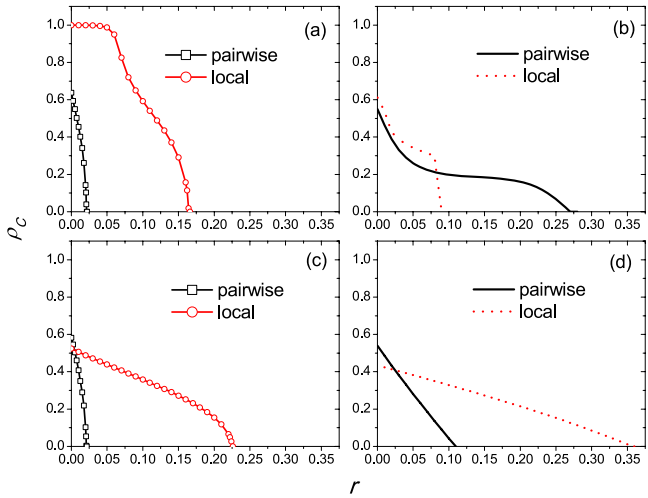}
\caption{Fraction of cooperators $\rho_C$ as a function of the cost-to-benefit ratio $r$, as obtained for $K=0.1$ [panels (a) and (b)] and $K=0.83$ [panels (c) and (d)]. Results presented in panels (a) and (c) were obtained by means of Monte Carlo simulations, while those presented in
panels (b) and (d) were obtained by means of pair approximation (see Methods section for details). Figure legend indicates whether pairwise or locally influenced strategy updating was used.}
\label{base}
\end{figure}

We begin by presenting the fraction of cooperators $\rho_C$ as a function of the cost-to-benefit ratio $r=c/b$ at two temperatures, namely at $K=0.1$ and $K=0.83$. The usage of the latter value is motivated by recent empirical research from behavioral science \cite{traulsen_pnas10}, although essentially, as we will show in what follows, the temperature, i.e., the level of uncertainty by strategy adoptions, does not play a decisive role. Results for both the pairwise and locally influenced strategy updating are presented in Fig.~\ref{base}(a,c). It can be observed that for $K=0.1$ the evolution of cooperation is promoted across the whole applicable span of $r$ if the traditionally used pairwise strategy updating is replaced by the proposed local influence based strategy updating. For $K=0.83$, however, the outcome is a bit less clear-cut. While pairwise imitation fails to sustain cooperative behavior at such high values of $r$ as locally influenced strategy updating, it is nevertheless more apt for achieving complete cooperator dominance. As we will show in what follows, it is indeed the case that locally influenced strategy updating often fails to completely eliminate defectors at small values of $r$, yet it opens up the possibility of survival of cooperators even under harsh defector-friendly conditions.

\begin{figure}
\centering
\includegraphics[width=8.5cm]{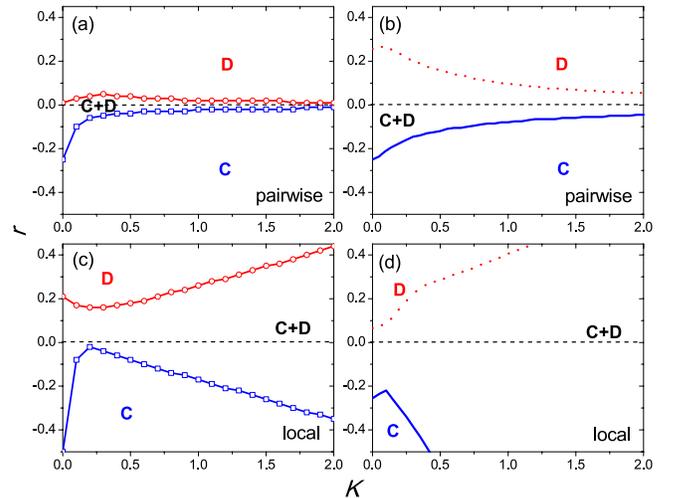}
\caption{Full $K-r$ phase diagrams, as obtained by means of Monte Carlo simulations [panels (a) and (c)] and by means of pair approximation [panels (b) and (d)]. Upper red (lower blue) lines
denote the boundaries between the mixed $C+D$ and homogeneous $D$ $(C)$ phases.}
\label{phase}
\end{figure}

These simulation results can be corroborated by results of pair approximations (see Methods for details), which we present in Fig.~\ref{base}(b,d). The general trends are predicted correctly, although as expect, the beneficial effect of network reciprocity \cite{nowak_n92b} at low values of $r$ are underestimated. It is worth mentioning that the pair approximation is in general more accurate for larger values of $K$ \cite{szabo_pre05}, and indeed it can be observed that the agreement with simulation results is better for $K=0.83$ than it is for $K=0.1$. In particular, for $K=0.83$ the pair approximation method correctly predicts the occurrence of an intersection point [compare panels (c) and (d)]. Altogether, results of pair approximations corroborate the conclusion that the survivability of cooperators, especially at high values of $r$, is substantially promoted by locally influenced strategy updating.

Further adding to the robustness of this conclusion are results presented in Fig.~\ref{phase}(a,c), where we present full $K-r$ phase diagrams for both considered updating rules. It can be observed that the positive impact of local influence on the evolution of cooperation persists across large regions of $K$. On the other hand, the presented phase diagrams also evidence more clearly the failure of the proposed updating rule to lead to an absorbing $C$ phase. Moreover, there is a notable qualitative difference in the critical behavior that is evoked by the updating rule. By focusing on the $D \to C+D$ phase boundaries, it can be observed that for pairwise strategy updating there exists an optimal value of $K$ at which cooperators thrive best. Note that the $D \to C + D$ phase boundary is bell-shaped, indicating that $K \approx 0.3$ is the optimal temperature at which cooperators are able to survive at the highest value of $r$. For strategy updating based on local influence, however, this feature is absent. The $D \to C + D$ phase boundary is in fact an inverted bell, indicating the existence of the worst rather than an optimal value of $K$. Notably, the results for pairwise strategy updating are in agreement with previous works \cite{szabo_pre05, perc_njp06a, vukov_pre06}, where it was shown that the lack of overlapping triangles, as is the case for the square lattice as well as for random regular graphs, introduces an optimal uncertainty for the evolution of cooperation. Conversely, the results obtained by considering local influence suggest that the system is behaving as if overlapping triangles were in fact present in the interaction network. Note that in the latter case an optimal $K$ for the evolution of cooperation does not exist. This leads us to the conclusion that the interaction network is effectively altered when the local influence is taken into account. In particular, triplets of players that are not connected by means of the original interaction graph (the square lattice) become effectively connected through the joint participation of players in the same local groups (neighborhoods) that are subject to the same local influence. An identical effect was indeed observed by the study of the public goods game \cite{szolnoki_pre09c}, where triplets also became effectively connected because of the participation of players in the same groups. Below, we will provide further evidence concerning the effective linkage of triples of players, which is essentially a side effect of locally influenced strategy updating. Another interesting observation is that the parameter region of the mixed $C+D$ phase in general widens as $K$ increases, which is in contract to the results obtained by means of pairwise strategy updating.

\begin{figure}
\centering
\includegraphics[width=8.5cm]{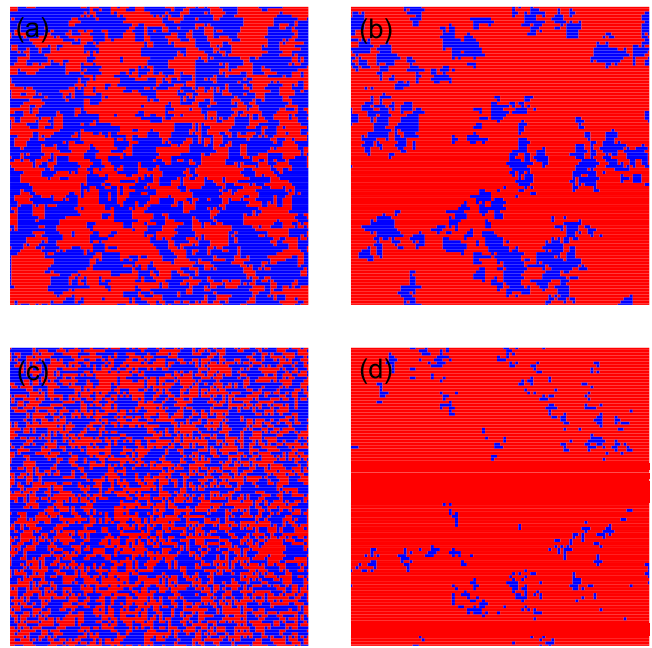}
\caption{Characteristic snapshots of spatial patterns formed by cooperators (blue) and defectors (red) under pairwise imitation [(a) $r=0.004$, (b) $r=0.019$] and under strategy updating based on local influence [(c) $r=0.004$, (d) $r=0.221$]. The size of the square lattice was $100 \times 100$ and $K=0.83$. (a) In this snapshot there are $77$ clusters, ranging in size from a single cooperator to $3042$ cooperators, with a weighted average size of $1925.21$. The stationary fraction of cooperators is $\rho_C \approx 0.52$. (b) In this snapshot there are $99$ clusters, ranging in size from a single cooperator to $162$ cooperators, with a weighted average size of $70.01$. The stationary fraction of cooperators is $\rho_C \approx 0.19$. These characteristics are significantly different in the bottom two snapshots. (c) In this snapshot there are $439$ clusters, ranging in size from a single cooperator to $427$ cooperators, with a weighted average size of $137.69$. The stationary fraction of cooperators is $\rho_C \approx 0.52$. (d) In this snapshot there are $164$ clusters, ranging in size from a single cooperator to $19$ cooperators, with a weighted average size of $6.63$. The stationary fraction of cooperators is $\rho_C \approx 0.05$. Note that in snapshots (a) and (c) the densities of cooperators for both update rules are practically identical, while nearer to the extinction thresholds [panels (b) and (d)] they differ quite significantly.}
\label{snapshot}
\end{figure}

We have also constructed full $K-r$ phase diagrams by means of pair approximations. Figure~\ref{phase}(b,d) features the obtained results, from which it follows that qualitative features, compared to the simulation results, are again captured fairly accurately, although the extent of the parameter region of the mixed $C+D$ phase is overestimated. Expectedly, the predictions are also less accurate near the phase boundaries, which is because the pair approximation does not take into account loops nor does it take into account long-range correlations, which however, have a noticeable effect especially in the vicinity of critical transitions \cite{hauert_ajp05}.

In order to obtain an understanding of the reported observations, we proceed with the presentation of characteristic spatial patterns, as obtained for both pairwise and locally influenced strategy updating, in Fig.~\ref{snapshot}. Regardless of which update rule is applied, cooperators form compact clusters by means of which they are able to exploit the mechanism of network reciprocity \cite{nowak_n92b}. If the value of $r$ is small, the clusters are larger and more compact than for higher values of $r$. On the other hand, the spatial patterns emerging under the two update rules also have noticeable dissimilarities. Foremost, given a value of $r$, pairwise strategy updating yields larger clusters than locally influenced strategy updating, even if the density of cooperators is approximately the same [compare panels (a) and (c)]. Nearer to the extinction threshold the stationary densities differ, yet the difference in the spatial patterns the two rules generate becomes most apparent [compare panels (b) and (d)].

\begin{figure}
\centering
\includegraphics[width=6cm]{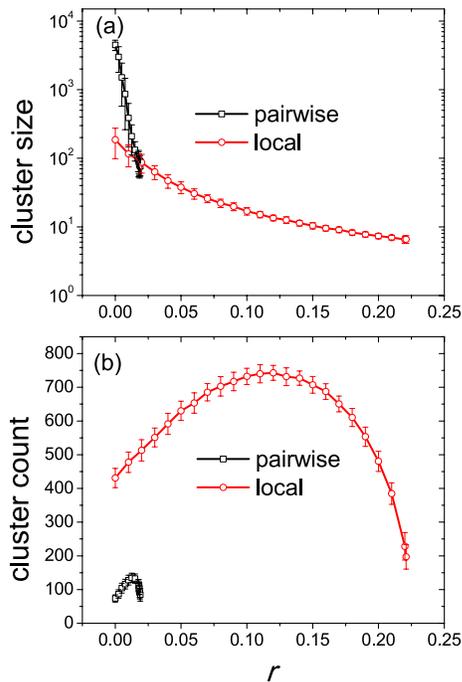}
\caption{Macroscopic properties of cooperative clusters in the dependence on the cost-to-benefit ratio $r$. Cluster size (a) and cluster count (b) are depicted for pairwise and locally influenced strategy updating. In both cases the cluster size decreases as $r$ increases, while the cluster count reaches a maximum at a certain value of $r$ and then decreases. Note that for pairwise imitation a minimum cluster size of about $76.18$ is required for cooperators to survive. Taking into account the local influence of the neighbors reduces this to $6.61$. The depicted results were determined in the stationary state on $100 \times 100$ square lattices and by using $K=0.83$. Error bars indicate the standard deviation.}
\label{cluster}
\end{figure}

The visual inspection of the characteristic spatial patterns invites a quantitative analysis of the exposed differences, the results of which are presented in Fig.~\ref{cluster} separately for both updating rules. It can be observed that, in general, as $r$ increases, the cluster size decreases. The number of clusters, on the other hand, is maximal at an intermediate value of $r$. Concrete $r$ values, however, differ significantly for the two considered strategy updating rules. In particular, by pairwise strategy updating both the clusters size and the number of clusters are shifted significantly towards lower values of $r$. One reason is obviously that pairwise strategy updating simply does not support the survivability of cooperators by as high values of $r$ as locally influenced strategy updating. Nonetheless, the fact that for any given value of $r$, where comparisons are possible, the typical cluster size obtained with pairwise strategy updating is much larger than the one obtained with locally influenced strategy updating begets the conclusion that there are significant differences in the way cooperators cluster to withstand being wiped out by defectors. Note that for cooperators to survive under pairwise updating the minimally required cluster size is $\approx 76.18$, while for locally influenced updating it is only $6.61$. Moreover, for pairwise strategy updating the cluster size decreases much faster, which speaks in favor of the increased stability of the clusters under locally influenced strategy updating.

\begin{figure}
\centering
\includegraphics[width=8.5cm]{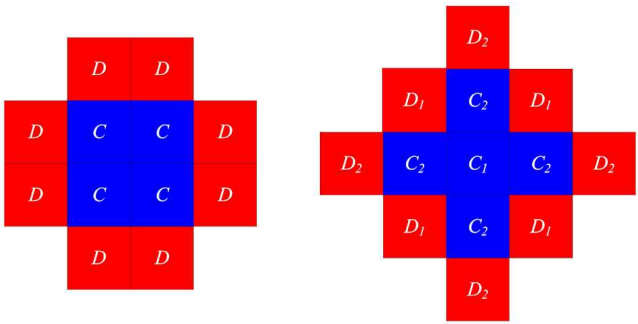}
\caption{Schematic presentation of two representative cooperative (blue) clusters surrounded by defectors (red). The cluster depicted left has no chances of survival under pairwise or locally influenced strategy updating. The cluster on the right, however, cannot prevail under pairwise imitation, but can do so under locally influenced strategy updating. This is because the core of the cooperative cluster ($C_1$ in the figure) is quarantined from defectors in case imitation proceeds according to local influence (see main text for details).}
\label{pattern}
\end{figure}

To confirm these conjectures, we present in Fig.~\ref{pattern} two typical $C$-cluster configurations and analyze the survivability of cooperators separately for each particular case. For the sake of simplicity but without loss of generality, we consider for the following analysis only the $K \to 0$ limit. Then if the payoff of each cooperator along the boundary is
larger than that of each defector in its neighborhood, we are allowed to conclude that such a $C$-cluster will survive. For the left $C$-cluster pattern in Fig.~\ref{pattern} under pairwise updating, the payoffs of a cooperator $C$ ($P_C$) and defector $D$ ($P_D$) along the boundary are
\begin{equation}
\label{eq.14}{P_C} = 2{\kern 1pt} {\kern 1pt} {\kern 1pt}
{\rm{and}}{\kern 1pt} {\kern 1pt} {\kern 1pt} {P_D} = 1 + 4r,
\end{equation}
respectively. For locally influenced updating, however, the average payoff of cooperators
(${\bar P}_C$) and the average payoff of defectors (${\bar P}_D$) along the boundary are given by
\begin{equation}
\label{eq.15} {{\bar P}_C} = 2{\kern 1pt} {\kern 1pt} {\kern 1pt}
{\rm{and}}{\kern 1pt} {\kern 1pt} {\kern 1pt} {{\bar P}_D} = 1 +
4r,
\end{equation}
respectively. Thus for such a $C$-cluster pattern to survive, both update rules lead to $r <  - 0.25$. Indeed, neither locally influenced nor pairwise strategy updating support the survivability of such a pattern. Performing the same analysis for the configuration on the right, however, yields a different outcome. The payoff of a cooperator $C_2$ ($P_{C_2}$) on the boundary and that of the two types of defectors $D_1$ and $D_2$ ($P_{D_2}$ and $P_{D_1}$) are
\begin{equation}
\label{eq.16} {P_{{C_2}}} = 1,{\kern 1pt} {\kern 1pt} {\kern 1pt}
{P_{{D_1}}} = 2 + 4r{\kern 1pt} {\kern 1pt} {\kern 1pt}
{\rm{and}}{\kern 1pt} {\kern 1pt} {P_{{D_2}}} = 1 + 4r,
\end{equation}
respectively. For locally influenced updating the corresponding payoffs are
\begin{equation}
\label{eq.17} {{\bar P}_C} = \frac{5}{2}{\kern 1pt} {\kern 1pt}
{\kern 1pt} {\rm{and}}{\kern 1pt} {\kern 1pt} {\kern 1pt} {{\bar
P}_D} = \frac{5}{3} + 4r.
\end{equation}
Accordingly, we find that under pairwise updating the condition for survivability is $r<-0.25$, while under locally influenced updating it is only $r < \frac{5}{{24}}$. Hence, locally influenced strategy updating can warrant the survivability of cooperators when grouped in this way, while pairwise updating can not. Note also that the $C$-cluster configuration on the right of Fig.~\ref{pattern} is the smallest one which can persist in the population under the most hostile conditions under locally influenced strategy updating. Based on this analysis, we can in fact estimate the extinction threshold $r = \frac{5}{{24}} \approx 0.21$ in the limit $K \to 0$, and indeed we find excellent agreement between this analytical approximation and the simulation results presented in Fig.~\ref{phase}(c).

With these insights, we argue that local influence based strategy updating can support the survivability of cooperative behavior only if the core of the $C$-cluster is isolated from defectors (compare left and right configuration of Fig.~\ref{pattern}), because cooperators along the boundary can then gain a higher level of support from the cluster and thus protect themselves against being exploited by defectors. In previous works, where only pairwise strategy updating was considered, individual players were concerned only with their own payoffs when updating their strategies. However, if individuals are exposed to the local influence, i.e., they care about the average performance of the strategies in their neighborhood, cooperators can benefit not only from their own payoffs, but also from the payoffs of their cooperative neighbors. In this sense, locally influenced strategy updating further strengthens the linkage between cooperators within cooperative clusters, and so cooperators can reciprocate with each other on a profounder and altogether more effective level.

In terms of the robustness of the described mechanism to variations of the interaction network, our preliminary investigations indicate that cooperation is always promoted on regular small-world networks with different rewiring probabilities \cite{szabo_jpa04}, as well as on scale-free networks \cite{barabasi_s99} provided the payoffs are normalized with the number of neighbors \cite{szolnoki_pa08}. If the payoffs on scale-free networks are not normalized with the number of neighbors \cite{santos_prl05}, the promotion of cooperation due to local influence, compared to the traditional pairwise strategy updating, may be compromised. Additional research is needed, however, to clarify conclusively the potential negative impact of strongly heterogeneous degree distributions on the newly identified mechanism for the promotion of cooperation. It would also be of much interest to clarify the role of zero-determinant strategies \cite{press_pnas12, stewart_pnas12}, which point to major paradigm shifts in the resolution of social dilemmas.

\section*{Discussion}
Summarizing, we have analyzed the impact of ``local influence'' on the evolution of cooperation in the spatial prisoner's dilemma game. Instead of the performance of a single neighbor, players considered the average performance of the two strategies within their neighborhoods. We have shown that by going beyond the traditionally assumed pairwise strategy updating, the evolution of cooperation can be promoted. We have determined full $K-r$ phase diagrams by means of simulations and pair approximation methods, which both indicate that this effect is robust against uncertainty by strategy adoptions. Moreover, the phase separation lines indicate that the consideration of local influence effectively changes the interaction network as an optimal $K$ is no longer inferable. This is characteristic for interaction networks with overlapping triangles \cite{szabo_pre05, vukov_pre06}, which are obviously not part of the square lattice topology that we have employed. By analyzing the macroscopic features of emerging spatial patterns as well as the survivability of typical cooperative clusters, we have provided further insights as to how the consideration of local influence changes the evolutionary dynamics.

Lastly, it is worth relating the presently considered strategy updating rule to previous game-theoretical models. By the win-stay-lose-shift rule \cite{nowak_n93, macy_pnas02, chen_xj_pre08, liu_yk_epl11, liu_yk_pone12}, for example, each individual has an aspiration according to which it judges whether or not to change strategy. The aspiration, however, is traditionally assumed to be constant. In our case, on the other hand, we relax this assumption by considering the aspiration as a dynamical quantity. Note that the average payoff of the strategy that is not adopted by the focal player can in fact be regarded as the aspiration level. This in turn implies that here the aspiration depends on the outcome of the game, and hence is subject to change. Moreover, the present rule can be regarded as a learning rule. The difference from the traditional single role model learning rule is that in the present case the strategy update depends not on the comparison of a pair of individuals, but on the comparison of two groups of individuals, each involving several individuals adopting the same strategy. Overall, we hope that these considerations, and in particular the consideration of local influence and the ``wisdom of crowds'' \cite{surowiecki_04, lorenz_pnas11, szolnoki_srep12}, will motivate further research aimed at promoting our understanding of the evolution of cooperation.

\section*{Methods}
\subsection*{Mathematical model}
Players are located on the vertices of a $L\times L$ square lattice with periodic
boundary conditions. Each individual is initially designated either as a cooperator $C$ or defector $D$ with equal probability. For the pairwise imitation strategy updating rule \cite{szabo_pre98} (we use the label ``pairwise'' in the figure legends when applying this rule), Monte
Carlo simulations of the game are carried out comprising the following elementary steps. First, a randomly selected player $x$ collects its payoff $P_x$ by interacting with its four nearest
neighbors. For the purpose of payoff evaluation, it is worth introducing unit vectors $S = {[1,0]^T}$ and ${[0,1]^T}$ for cooperators and defectors, respectively. The payoff matrix is
\[M = \left[ {\begin{array}{*{20}{c}}
   1 & 0  \\
   {1 + r} & r  \\
\end{array}} \right],\]
where $r \in (0, 1)$ is the cost-to-benefit ratio. The payoff of player $x$ is thus
\begin{equation}
P_x = \sum\limits_{z \in \Gamma (x)} {S_x^TM{S_z}},
\end{equation}
where ${\Gamma (x)}$ represents its neighborhood. Subsequently, one randomly chosen neighbor $y$ of player $x$ also acquires its
payoff $P_y$ identically as previously player $x$.

After the evaluation of payoffs, players consider changing their strategies. In particular, player $x$ adopts the strategy $S_{y}$ of the randomly selected neighbor with the probability
\begin{equation}
\label{eq.1}T({P_y} - {P_x}) = \frac{1}{{1 + \exp [({P_x} - {P_y})/K]}},
\end{equation}
where $K$ is the uncertainty by strategy adoptions \cite{szabo_pr07}. If the local influence is taken into account (we use the label ``local'' in the figure legends when applying this rule), however, the elementary steps are as follows. First, we randomly choose a player $x$ with the strategy $S_{x}$. Next, we evaluate the average payoff $\bar P_{S_x}$ of those players who adopt the same strategy $S_{x}$, as well as the average payoff $\bar P_{\bar {S_x}}$ of those players who adopt the opposite strategy $\bar {S_{x}}$ of player $x$, if any, within the neighborhood. In particular, we have
\begin{equation}
{\bar P_{{S_x}}} = \frac{{\sum\limits_{z \in \Gamma (x)} {{P_z}\delta (\bar S_x^T{S_z})}  + {P_x}}}{{\sum\limits_{z \in \Gamma (x)} {\delta (\bar S_x^T{S_z})}  + 1}}
\end{equation}
and
\begin{equation}
{\bar P_{{{\bar S}_x}}} = \frac{{\sum\limits_{z \in \Gamma (x)} {{P_z}\delta (S_x^T{S_Z})} }}{{\sum\limits_{z \in \Gamma (x)} {\delta (S_x^T{S_Z})} }},
\end{equation}
where the Dirac delta function $\delta (x)$ satisfies
\[\delta (x) = \left\{ {\begin{array}{*{20}{c}}
   {0,{\kern 1pt} {\kern 1pt} {\kern 1pt} {\rm{if}}{\kern 1pt} x \ne 0}  \\
   {1,{\kern 1pt} {\kern 1pt} {\kern 1pt} {\rm{if}}{\kern 1pt} x = 0}  \\
\end{array}} \right..\]
Lastly, player $x$ will adopt the strategy $\bar {S_{x}}$ with the probability
\begin{equation}
\label{eq.2}T({{\bar P}_{{{\bar S}_x}}} - {{\bar P}_{{S_x}}}) =
\frac{1}{{1 + \exp [ - ({{\bar P}_{{{\bar S}_x}}} - {{\bar
P}_{{S_x}}})/K]}},
\end{equation}
where $K$ is, as by pairwise imitation, the uncertainty by strategy adoptions.

The presented simulation results were obtained by using $L=100-400$ depending on the proximity to phase separation lines and the size of the emerging spatial patterns. In accordance with the random sequential update, each full Monte Carlo step, which consists of repeating the elementary steps $L\times L$ times corresponding to all players, gives a chance once on average for every player to alter its strategy. The stationary frequency of cooperators $\rho _{C}$ is determined by averaging over ${10^4}$ Monte Carlo steps in the stationary state after sufficiently long relaxation times. In general, the stationary state has been considered to be reached when the average of the
cooperation level becomes time-independent. To further increase the accuracy of our simulations, we have averaged the final outcome over $50$ independent initial conditions.

\subsection*{Pair approximations}
Let $p_C$ and $p_D=1-p_C$ denote the frequencies of cooperators and defectors, respectively, and let $p_{CC}$, $p_{CD}$, $p_{DC}$ and $p_{DD}$ represent the frequencies of $CC$, $CD$, $DC$ and $DD$ pairs, respectively. Then $q_{X|Y}=p_{XY}/p_Y$ with $X, Y \in {C, D}$ specifies the conditional probability to find an $X$-player given that the neighboring node is occupied by
an $Y$-player. Note that here $X$, $Y$ and $Z$ denote either $C$ or $D$. Instead of the first-order approximation considering the frequency of strategies as in the well-mixed population, the pair approximation tracks the frequencies of strategy pairs $p_{XY}$ ($X, Y \in {C, D}$). The probabilities of larger configurations are approximated by the frequencies of configurations not more complex than pairs. Based on the compatibility condition ${p_X} = \sum\nolimits_Y {{p_{XY}}}$, the symmetry condition ${p_{XY}}= {p_{YX}}$, and closure conditions, $p_C$ and $p_{CC}$ can fully determine the dynamics of the system. While the pair approximation for pairwise imitation is well-known and can be looked up for example in the Appendix of \cite{szabo_pr07} or more recently \cite{fu_jtb10}, for the imitation based on local influence the derivations are as follows.

A defector is selected for strategy updating with the probability $p_D$. Let $k_C$ and $k_D$ denote the number of cooperators and defectors amongst the neighbors on a regular lattice with degree $k$, respectively. The frequency of such a configuration is
\begin{equation}
\frac{{k!}}{{{k_C}!{k_D}!}}q_{C|D}^{{k_C}}q_{D|D}^{{k_D}},
\end{equation}
and the payoff of the defector is ${P_D}({k_C},{k_D}) = (1 +
r) \cdot {k_C} + r \cdot {k_D}$. The configuration probability with which a neighboring cooperator has $k_{C}^{'}$ cooperators and $k_{D}^{'}$ defectors as its neighbors is
\begin{equation}
\frac{{(k - 1)!}}{{k_{C}^{'}!k_{D}^{'}!}}q_{C|CD}^{k_{C}^{'}}q_{D|CD}^{k_{D}^{'}},
\end{equation}
where $q_{X|YZ}$ gives the conditional probability that a player next to the $YZ$ pair is in state $X$. The payoff of the neighboring cooperator is ${P_C}(k_{C}^{'},k_{D}^{'}) = k_{C}^{'}$. Similarly, the configuration probability with which a neighboring defector has $k_{C}^{'}$ cooperators and $k_{D}^{'}$ defectors as its neighbors is
\begin{equation}
\frac{{(k - 1)!}}{{k_{C}^{'}!k_{D}^{'}!}}q_{C|DD}^{k_{C}^{'}}q_{D|DD}^{k_{D}^{'}},
\end{equation}
and the payoff of the neighboring defector is ${P_D}(k_{C}^{'},k_{D}^{'}) = (1 + r) \cdot k_{C}^{'} + r \cdot (k_{D}^{'} + 1)$. Thus, the average payoff ${\bar P}_C$ of cooperators that are neighbors of the focal defector is
\begin{equation}
\label{eq.3}\begin{array}{l}
 {{\bar P}_C} = \sum\limits_{k_{C}^{'} = 0}^{k - 1} {\frac{{(k - 1)!}}{{k_{C}^{'}!k_{D}^{'}!}}q_{C|CD}^{k_{C}^{'}}q_{D|CD}^{k_{D}^{'}} \cdot } {P_C}(k_{C}^{'},k_{D}^{'}) \\
 {\kern 1pt} {\kern 1pt} {\kern 1pt} {\kern 1pt} {\kern 1pt} {\kern 1pt} {\kern 1pt} {\kern 1pt} {\kern 1pt} {\kern 1pt} {\kern 1pt} {\kern 1pt} {\kern 1pt} {\kern 1pt} {\kern 1pt}  = (k - 1) \cdot {q_{C|CD}} .\\
 \end{array}
\end{equation}
The average payoff ${\bar P}_D$ of defectors that are neighbors of the focal defector, on the other hand, is
\begin{equation}
\label{eq.4}\begin{array}{l}
 {{\bar P}_D} = \frac{{{k_D} \cdot \sum\limits_{k_{C}^{'} = 0}^{k - 1} {\frac{{(k - 1)!}}{{k_{C}^{'}!k_{D}^{'}!}}q_{C|DD}^{k_{C}^{'}}q_{D|DD}^{k_{D}^{'}} \cdot } {P_D}(k_{C}^{'},k_{D}^{'}) + {P_D}({k_C},{k_D})}}{{{k_D} + 1}} \\
 {\kern 1pt} {\kern 1pt} {\kern 1pt} {\kern 1pt} {\kern 1pt} {\kern 1pt} {\kern 1pt} {\kern 1pt} {\kern 1pt} {\kern 1pt} {\kern 1pt} {\kern 1pt} {\kern 1pt} {\kern 1pt} {\kern 1pt}  = \frac{{{k_D} \cdot [(k - 1) \cdot {q_{C|DD}} + rk] + rk + {k_C}}}{{{k_D} + 1}}.
 \end{array}
\end{equation}
Consequently, $p_C$ increases by $1/N$ where $N=L^{2}$, with probability
\begin{equation}
\begin{array}{l}
\label{eq.5}\Pr {\rm{ob}}(\Delta {p_C} = \frac{1}{N}) = \\
 {p_D}
\cdot \sum\limits_{{k_C} = 1}^k
{\frac{{k!}}{{{k_C}!{k_D}!}}q_{C|D}^{{k_C}}q_{D|D}^{{k_D}} \cdot
T({{\bar P}_C} - {{\bar P}_D})},
 \end{array}
\end{equation}
where $T({{\bar P}_C} - {{\bar P}_D})$ is the individual transition probability given by Eq.~\ref{eq.2}. The number of $CC$ pairs increases by $k_C$, and thus $p_{CC}$ increases by $2k_C/(kN)$ with probability
\begin{equation}
\begin{array}{l}
\label{eq.6}\Pr {\rm{ob}}(\Delta {p_{CC}} = \frac{{2{k_C}}}{{kN}})
= \\ {p_D} \cdot
\frac{{k!}}{{{k_C}!{k_D}!}}q_{C|D}^{{k_C}}q_{D|D}^{{k_D}} \cdot
T({{\bar P}_C} - {{\bar P}_D}).
\end{array}
\end{equation}

A cooperator, on the other hand, is selected for strategy updating with the probability $p_C$.
The frequency of a configuration that there are $k_C$ cooperators and $k_D$ defectors
in the neighborhood of the focal cooperator is
\begin{equation}
\frac{{k!}}{{{k_C}!{k_D}!}}q_{C|C}^{{k_C}}q_{D|C}^{{k_D}},
\end{equation}
and the payoff of the focal cooperator is ${P_C}({k_C},{k_D}) =
{k_C}$. The configuration probability with which a neighboring cooperator
has $k_{C}^{'}$ cooperators and $k_{D}^{'}$ defectors as its
neighbors is 
\begin{equation}
\frac{{(k - 1)!}}{{k_{C}^{'}!k_{D}^{'}!}}q_{C|CC}^{k_{C}^{'}}q_{D|CC}^{k_{D}^{'}},
\end{equation}
and the payoff of the neighboring cooperator is
${P_C}(k_{C}^{'},k_{D}^{'}) = k_{C}^{'} + 1$. Similarly, the configuration probability with which a neighboring defector has $k_{C}^{'}$ cooperators and $k_{D}^{'}$ defectors as its neighbors is
\begin{equation}
\frac{{(k - 1)!}}{{k_{C}^{'}!k_{D}^{'}!}}q_{C|DC}^{k_{C}^{'}}q_{D|DC}^{k_{D}^{'}},
\end{equation}
and the payoff of the neighboring defector is ${P_D}(k_{C}^{'},k_{D}^{'}) = (1 + r) \cdot (k_{C}^{'} + 1) +rk_{D}^{'}$. Thus the average payoff ${\bar P}_C$ of cooperators in the neighborhood of the focal cooperator is
\begin{equation}
\label{eq.7}\begin{array}{l}
 {{\bar P}_C} = \frac{{{k_C} \cdot \sum\limits_{k_{C}^{'} = 0}^{k - 1} {\frac{{(k - 1)!}}{{k_{C}^{'}!k_{D}^{'}!}}q_{C|CC}^{k_{C}^{'}}q_{D|CC}^{k_{D}^{'}} \cdot } {P_C}(k_{C}^{'},k_{D}^{'}) + {P_C}({k_C},{k_D})}}{{{k_C} + 1}} \\
 {\kern 1pt} {\kern 1pt} {\kern 1pt} {\kern 1pt} {\kern 1pt} {\kern 1pt} {\kern 1pt} {\kern 1pt} {\kern 1pt} {\kern 1pt} {\kern 1pt} {\kern 1pt} {\kern 1pt} {\kern 1pt} {\kern 1pt}  = \frac{{{k_C} \cdot [(k - 1) \cdot {q_{C|CC}} + 2]}}{{{k_C} + 1}} ,\\
 \end{array}
\end{equation}
while, the average payoff ${\bar P}_D$ of defectors in the neighborhood of the focal cooperator is
\begin{equation}
\label{eq.8}\begin{array}{l}
 {{\bar P}_D} = \sum\limits_{k_{C}^{'} = 0}^{k - 1} {\frac{{(k - 1)!}}{{k_{C}^{'}!k_{D}^{'}!}}q_{C|DC}^{k_{C}^{'}}q_{D|DC}^{k_{D}^{'}} \cdot } {P_D}(k_{C}^{'},k_{D}^{'}) \\
 {\kern 1pt} {\kern 1pt} {\kern 1pt} {\kern 1pt} {\kern 1pt} {\kern 1pt} {\kern 1pt} {\kern 1pt} {\kern 1pt} {\kern 1pt} {\kern 1pt} {\kern 1pt} {\kern 1pt} {\kern 1pt}  = (k - 1) \cdot {q_{C|DC}} + 1 + rk .\\
 \end{array}
\end{equation}
Thus $p_C$ decreases by $1/N$ with probability
\begin{equation}
\begin{array}{l}
\label{eq.9}\Pr {\rm{ob}}(\Delta {p_C} =  - \frac{1}{N}) = \\{p_C}
\cdot \sum\limits_{{k_C} = 0}^{k - 1}
{\frac{{k!}}{{{k_C}!{k_D}!}}q_{C|C}^{{k_C}}q_{D|C}^{{k_D}} \cdot
T({{\bar P}_D} - {{\bar P}_C})}.
\end{array}
\end{equation}
Moreover, the number of $CC$ pairs decreases by $k_C$ and thus $p_{CC}$ decreases by $2k_C/(kN)$ with probability
\begin{equation}
\begin{array}{l}
\label{eq.10}\Pr {\rm{ob}}(\Delta {p_{CC}} =  -
\frac{{2{k_C}}}{{kN}}) = \\ {p_C} \cdot
\frac{{k!}}{{{k_C}!{k_D}!}}q_{C|C}^{{k_C}}q_{D|C}^{{k_D}} \cdot
T({{\bar P}_D} - {{\bar P}_C}).
\end{array}
\end{equation}

These derivations lead us to the master equations
\begin{widetext}
\begin{equation}
\label{eq.11}{{\dot p}_C} = \Pr {\rm{ob}}(\Delta {p_C} =
\frac{1}{N}) - \Pr {\rm{ob}}(\Delta {p_C} =  - \frac{1}{N}) \ \ \ \ {\rm and}
\end{equation}
\begin{equation}
\label{eq.12}{{\dot p}_{CC}} = \sum\limits_{{k_C} = 0}^k
{\frac{{2{k_C}}}{k}[\Pr {\rm{ob}}(\Delta {p_{CC}} =
\frac{{2{k_C}}}{{kN}}) - \Pr {\rm{ob}}(\Delta {p_{CC}} =  -
\frac{{2{k_C}}}{{kN}})]}.
\end{equation}
\end{widetext}
Although these equations are per derivation exact, they do depend on the density of triplet configurations which are outside their scope. Thus, in order to ``close" the system of
differential equations, the triplet configuration probabilities have to be approximated by probabilities of configurations that are not more complex than pairs. Note that by using different closure conditions, we can in general obtain different pair approximations. Here we employ the so-called ordinary pair approximation method, where only the first-order pair
correlations are considered. We thus have ${q_{X|YZ}} \approx {q_{X|Y}}$.\\ \\

\noindent \textbf{Acknowledgments} \\
This research was supported by the National 973 Program (grant 2012CB821203), the National Natural Science Foundation of China (grants 61020106005, 10972002 and 61203374), and the Slovenian Research Agency (grant J1-4055).

\noindent \\ \textbf{Author contributions} \\
Xiaofeng Wang, Matja\v{z} Perc, Yongkui Liu, Xiaojie Chen and Long Wang designed and performed this research.


\clearpage

\begin{thebibliography}{78}
\expandafter\ifx\csname natexlab\endcsname\relax\def\natexlab#1{#1}\fi
\expandafter\ifx\csname url\endcsname\relax
  \def\url#1{\texttt{#1}}\fi
\expandafter\ifx\csname urlprefix\endcsname\relax\def\urlprefix{URL }\fi

\bibitem[{Axelrod(1984)}]{axelrod_84}
Axelrod, R.
\newblock \emph{The Evolution of Cooperation} (Basic Books, New York, 1984).

\bibitem[{Bowles \& Gintis(2011)}]{bowles_11}
Bowles, S. \& Gintis, H.
\newblock \emph{A Cooperative Species: Human Reciprocity and Its Evolution}
  (Princeton Univ. Press, Princeton, NJ, 2011).

\bibitem[{Hrdy(2011)}]{hrdy_11}
Hrdy, S.~B.
\newblock \emph{Mothers and Others: The Evolutionary Origins of Mutual
  Understanding} (Harvard Univ. Press, Cambridge, Massachusetts, 2011).

\bibitem[{Nowak \& Highfield(2011)}]{nowak_11}
Nowak, M.~A. \& Highfield, R.
\newblock \emph{SuperCooperators: Altruism, Evolution, and Why We Need Each
  Other to Succeed} (Free Press, New York, 2011).

\bibitem[{Doebeli \& Hauert(2005)}]{doebeli_el05}
Doebeli, M. \& Hauert, C.
\newblock Models of cooperation based on Prisoner's Dilemma and Snowdrift game.
\newblock \emph{Ecol. Lett.} \textbf{8}, 748--766 (2005).

\bibitem[{Nowak(2006{\natexlab{a}})}]{nowak_s06}
Nowak, M.~A.
\newblock Five Rules for the Evolution of Cooperation.
\newblock \emph{Science} \textbf{314}, 1560--1563 (2006{\natexlab{a}}).

\bibitem[{Szab{\'o} \& F{\'a}th(2007)}]{szabo_pr07}
Szab{\'o}, G. \& F{\'a}th, G.
\newblock Evolutionary games on graphs.
\newblock \emph{Phys. Rep.} \textbf{446}, 97--216 (2007).

\bibitem[{Schuster \emph{et~al.}(2008)Schuster, Kreft, Schroeter \&
  Pfeiffer}]{schuster_jbp08}
Schuster, S., Kreft, J.-U., Schroeter, A. \& Pfeiffer, T.
\newblock Use of Game-Theoretical Methods in Biochemistry and Biophysics.
\newblock \emph{J. Biol. Phys.} \textbf{34}, 1--17 (2008).

\bibitem[{Roca \emph{et~al.}(2009)Roca, Cuesta \& S{\'a}nchez}]{roca_plr09}
Roca, C.~P., Cuesta, J.~A. \& S{\'a}nchez, A.
\newblock Evolutionary game theory: Temporal and spatial effects beyond
  replicator dynamics.
\newblock \emph{Phys. Life Rev.} \textbf{6}, 208--249 (2009).

\bibitem[{Perc \& Szolnoki(2010)}]{perc_bs10}
Perc, M. \& Szolnoki, A.
\newblock Coevolutionary games -- a mini review.
\newblock \emph{BioSystems} \textbf{99}, 109--125 (2010).

\bibitem[{Hofbauer \& Sigmund(1998)}]{hofbauer_98}
Hofbauer, J. \& Sigmund, K.
\newblock \emph{Evolutionary Games and Population Dynamics} (Cambridge Univ.
  Press, Cambridge, UK, 1998).

\bibitem[{Nowak(2006{\natexlab{b}})}]{nowak_06}
Nowak, M.~A.
\newblock \emph{Evolutionary Dynamics} (Harvard Univ. Press, Cambridge, MA,
  2006{\natexlab{b}}).

\bibitem[{Sigmund(2010)}]{sigmund_10}
Sigmund, K.
\newblock \emph{The Calculus of Selfishness} (Princeton Univ. Press, Princeton,
  MA, 2010).

\bibitem[{Glance \& Huberman(1994)}]{glance_sa94}
Glance, N.~S. \& Huberman, B.~A.
\newblock The Dynamics of Social Dilemmas.
\newblock \emph{Scientific American} 76--81 (1994).

\bibitem[{Hamilton(1964)}]{hamilton_wd_jtb64b}
Hamilton, W.~D.
\newblock Genetical evolution of social behavior \protect{II}.
\newblock \emph{J. Theor. Biol.} \textbf{7}, 17--52 (1964).

\bibitem[{Trivers(1971)}]{trivers_qrb71}
Trivers, R.~L.
\newblock The evolution of reciprocal altruism.
\newblock \emph{Q. Rev. Biol.} \textbf{46}, 35--57 (1971).

\bibitem[{Nowak \& Sigmund(1998)}]{nowak_n98}
Nowak, M.~A. \& Sigmund, K.
\newblock Evolution of indirect reciprocity by image scoring.
\newblock \emph{Nature} \textbf{393}, 573--577 (1998).

\bibitem[{Wilson(1975)}]{wilson_ds_pnas75}
Wilson, D.~S.
\newblock A Theory of Group Selection.
\newblock \emph{Proc. Nat. Acad. Sci. USA} \textbf{72}, 143--146 (1975).

\bibitem[{Nowak \& May(1992)}]{nowak_n92b}
Nowak, M.~A. \& May, R.~M.
\newblock Evolutionary Games and Spatial Chaos.
\newblock \emph{Nature} \textbf{359}, 826--829 (1992).

\bibitem[{Santos \& Pacheco(2005)}]{santos_prl05}
Santos, F.~C. \& Pacheco, J.~M.
\newblock Scale-free networks provide a unifying framework for the emergence of
  cooperation.
\newblock \emph{Phys. Rev. Lett.} \textbf{95}, 098104 (2005).

\bibitem[{Santos \emph{et~al.}(2006{\natexlab{a}})Santos, Pacheco \&
  Lenaerts}]{santos_pnas06}
Santos, F.~C., Pacheco, J.~M. \& Lenaerts, T.
\newblock Evolutionary dynamics of social dilemmas in structured heterogeneous
  populations.
\newblock \emph{Proc. Natl. Acad. Sci. USA} \textbf{103}, 3490--3494
  (2006{\natexlab{a}}).

\bibitem[{Santos \emph{et~al.}(2006{\natexlab{b}})Santos, Rodrigues \&
  Pacheco}]{santos_prsb06}
Santos, F.~C., Rodrigues, J.~F. \& Pacheco, J.~M.
\newblock Graph topology plays a determinant role in the evolution of
  cooperation.
\newblock \emph{Proc. R. Soc. B} \textbf{273}, 51--55 (2006{\natexlab{b}}).

\bibitem[{G{\'o}mez-Garde{\~n}es \emph{et~al.}(2007)G{\'o}mez-Garde{\~n}es,
  Campillo, Moreno \& Flor{\' \i}a}]{gomez-gardenes_prl07}
G{\'o}mez-Garde{\~n}es, J., Campillo, M., Moreno, Y. \& Flor{\' \i}a, L.~M.
\newblock Dynamical Organization of Cooperation in Complex Networks.
\newblock \emph{Phys. Rev. Lett.} \textbf{98}, 108103 (2007).

\bibitem[{Poncela \emph{et~al.}(2007)Poncela, G{\'o}mez-Garde{\~n}es, Flor{\'
  \i}a \& Moreno}]{poncela_njp07}
Poncela, J., G{\'o}mez-Garde{\~n}es, J., Flor{\' \i}a, L.~M. \& Moreno, Y.
\newblock Robustness of cooperation in the evolutionary prisoner's dilemma on
  complex systems.
\newblock \emph{New J. Phys.} \textbf{9}, 184 (2007).

\bibitem[{Szolnoki \emph{et~al.}(2008{\natexlab{a}})Szolnoki, Perc \&
  Danku}]{szolnoki_pa08}
Szolnoki, A., Perc, M. \& Danku, Z.
\newblock Towards effective payoffs in the prisoner's dilemma game on
  scale-free networks.
\newblock \emph{Physica A} \textbf{387}, 2075--2082 (2008{\natexlab{a}}).

\bibitem[{Poncela \emph{et~al.}(2009{\natexlab{a}})Poncela,
  G{\'o}mez-Garde{\~n}es, Flor{\' \i}a, Moreno \& S{\'a}nchez}]{poncela_epl09}
Poncela, J., G{\'o}mez-Garde{\~n}es, J., Flor{\' \i}a, L.~M., Moreno, Y. \&
  S{\'a}nchez, A.
\newblock Cooperative scale-free networks despite the presence of defector
  hubs.
\newblock \emph{EPL} \textbf{88}, 38003 (2009{\natexlab{a}}).

\bibitem[{Perc(2009)}]{perc_njp09}
Perc, M.
\newblock Evolution of cooperation on scale-free networks subject to error and
  attack.
\newblock \emph{New J. Phys.} \textbf{11}, 033027 (2009).

\bibitem[{Santos \emph{et~al.}(2005)Santos, Rodrigues \&
  Pacheco}]{santos_pre05}
Santos, F.~C., Rodrigues, J.~F. \& Pacheco, J.~M.
\newblock Epidemic spreading and cooperation dynamics on homogeneous
  small-world networks.
\newblock \emph{Phys. Rev. E} \textbf{72}, 056128 (2005).

\bibitem[{Ren \emph{et~al.}(2007)Ren, Wang \& Qi}]{ren_pre07}
Ren, J., Wang, W.-X. \& Qi, F.
\newblock Randomness enhances cooperation: coherence resonance in evolutionary
  game.
\newblock \emph{Phys. Rev. E} \textbf{75}, 045101(R) (2007).

\bibitem[{Fu \emph{et~al.}(2007)Fu, Liu \& Wang}]{fu_epjb07}
Fu, F., Liu, L.-H. \& Wang, L.
\newblock Evolutionary prisoner's dilemma on heterogeneous
  \protect{Newman-Watts} small-world network.
\newblock \emph{Eur. Phys. J. B} \textbf{56}, 367--372 (2007).

\bibitem[{Perc(2006{\natexlab{a}})}]{perc_njp06c}
Perc, M.
\newblock Double resonance in cooperation induced by noise and network
  variation for an evolutionary prisoner's dilemma.
\newblock \emph{New J. Phys.} \textbf{8}, 183 (2006{\natexlab{a}}).

\bibitem[{Chen \& Wang(2008)}]{chen_xj_pre08}
Chen, X.-J. \& Wang, L.
\newblock Promotion of cooperation induced by appropriate payoff aspirations in
  a small-world networked game.
\newblock \emph{Phys. Rev. E} \textbf{77}, 017103 (2008).

\bibitem[{Vukov \& Szab{\'o}(2005)}]{vukov_pre05}
Vukov, J. \& Szab{\'o}, G.
\newblock Evolutionary prisoner's dilemma game on hierarchical lattices.
\newblock \emph{Phys. Rev. E} \textbf{71}, 036133 (2005).

\bibitem[{G{\'o}mez-Garde{\~n}es \emph{et~al.}(2008)G{\'o}mez-Garde{\~n}es,
  Poncela, Flor{\' \i}a \& Moreno}]{gomez-gardenes_jtb08}
G{\'o}mez-Garde{\~n}es, J., Poncela, J., Flor{\' \i}a, L.~M. \& Moreno, Y.
\newblock Natural Selection of Cooperation and Degree Hierarchy in
  Heterogeneous Populations.
\newblock \emph{J. Theor. Biol.} \textbf{253}, 296--301 (2008).

\bibitem[{Lee \emph{et~al.}(2011)Lee, Holme \& Wu}]{lee_s_prl11}
Lee, S., Holme, P. \& Wu, Z.-X.
\newblock Emergent Hierarchical Structures in Multiadaptive Games.
\newblock \emph{Phys. Rev. Lett.} \textbf{106}, 028702 (2011).

\bibitem[{Santos \emph{et~al.}(2008)Santos, Santos \& Pacheco}]{santos_n08}
Santos, F.~C., Santos, M.~D. \& Pacheco, J.~M.
\newblock Social diversity promotes the emergence of cooperation in public
  goods games.
\newblock \emph{Nature} \textbf{454}, 213--216 (2008).

\bibitem[{Perc \& Szolnoki(2008)}]{perc_pre08}
Perc, M. \& Szolnoki, A.
\newblock Social diversity and promotion of cooperation in the spatial
  prisoner's dilemma game.
\newblock \emph{Phys. Rev. E} \textbf{77}, 011904 (2008).

\bibitem[{Perc(2007)}]{perc_njp11}
Perc, M.
\newblock Does strong heterogeneity promote cooperation by group interactions?
\newblock \emph{New J. Phys.} \textbf{13}, 123027 (2007).

\bibitem[{Santos \emph{et~al.}(2012)Santos, Pinheiro, Lenaerts \&
  Pacheco}]{santos_jtb12}
Santos, F.~C., Pinheiro, F., Lenaerts, T. \& Pacheco, J.~M.
\newblock Role of diversity in the evolution of cooperation.
\newblock \emph{J. Theor. Biol.} \textbf{299}, 88--96 (2012).

\bibitem[{Zimmermann \emph{et~al.}(2004)Zimmermann, Egu{\'{\i}}luz \&
  Miguel}]{zimmermann_pre04}
Zimmermann, M.~G., Egu{\'{\i}}luz, V. \& Miguel, M.~S.
\newblock Coevolution of dynamical states and interactions in dynamic networks.
\newblock \emph{Phys. Rev. E} \textbf{69}, 065102(R) (2004).

\bibitem[{Zimmermann \& Egu{\'{\i}}luz(2005)}]{zimmermann_pre05}
Zimmermann, M.~G. \& Egu{\'{\i}}luz, V.
\newblock Cooperation, Social Networks and the Emergence of Leadership in a
  Prisoner's Dilemma with Local Interactions.
\newblock \emph{Phys. Rev. E} \textbf{72}, 056118 (2005).

\bibitem[{Pacheco \emph{et~al.}(2006{\natexlab{a}})Pacheco, Traulsen \&
  Nowak}]{pacheco_jtb06}
Pacheco, J.~M., Traulsen, A. \& Nowak, M.~A.
\newblock Active linking in evolutionary games.
\newblock \emph{J. Theor. Biol.} \textbf{243}, 437--443 (2006{\natexlab{a}}).

\bibitem[{Pacheco \emph{et~al.}(2006{\natexlab{b}})Pacheco, Traulsen \&
  Nowak}]{pacheco_prl06}
Pacheco, J.~M., Traulsen, A. \& Nowak, M.~A.
\newblock Coevolution of strategy and structure in complex networks with
  dynamical linking.
\newblock \emph{Phys. Rev. Lett.} \textbf{97}, 258103 (2006{\natexlab{b}}).

\bibitem[{Santos \emph{et~al.}(2006{\natexlab{c}})Santos, Pacheco \&
  Lenaerts}]{santos_ploscb06}
Santos, F.~C., Pacheco, J.~M. \& Lenaerts, T.
\newblock Cooperation prevails when individuals adjust their social ties.
\newblock \emph{PLoS Comput. Biol.} \textbf{2}, 1284--1290
  (2006{\natexlab{c}}).

\bibitem[{Fu \emph{et~al.}(2008)Fu, Hauert, Nowak \& Wang}]{fu_pre08b}
Fu, F., Hauert, C., Nowak, M.~A. \& Wang, L.
\newblock Reputation-based partner choice promotes cooperation in social
  networks.
\newblock \emph{Phys. Rev. E} \textbf{78}, 026117 (2008).

\bibitem[{Fu \emph{et~al.}(2009)Fu, Wu \& Wang}]{fu_pre09}
Fu, F., Wu, T. \& Wang, L.
\newblock Partner switching stabilizes cooperation in coevolutionary Prisoner's
  Dilemma.
\newblock \emph{Phys. Rev. E} \textbf{79}, 036101 (2009).

\bibitem[{Chen \emph{et~al.}(2009)Chen, Fu \& Wang}]{chen_xj_pre09b}
Chen, X., Fu, F. \& Wang, L.
\newblock Social tolerance allows cooperation to prevail in an adaptive
  environment.
\newblock \emph{Phys. Rev. E} \textbf{80}, 051104 (2009).

\bibitem[{Wu \emph{et~al.}(2009)Wu, Fu \& Wang}]{wu_t_epl09}
Wu, T., Fu, F. \& Wang, L.
\newblock Individual's expulsion to nasty environment promotes cooperation in
  public goods games.
\newblock \emph{EPL} \textbf{88}, 30011 (2009).

\bibitem[{Szolnoki \emph{et~al.}(2008{\natexlab{b}})Szolnoki, Perc \&
  Danku}]{szolnoki_epl08}
Szolnoki, A., Perc, M. \& Danku, Z.
\newblock Making new connections towards cooperation in the prisoner's dilemma
  game.
\newblock \emph{EPL} \textbf{84}, 50007 (2008{\natexlab{b}}).

\bibitem[{Poncela \emph{et~al.}(2008)Poncela, G{\'o}mez-Garde{\~n}es, Flor{\'
  \i}a, S\'anchez \& Moreno}]{poncela_ploso08}
Poncela, J., G{\'o}mez-Garde{\~n}es, J., Flor{\' \i}a, L.~M., S\'anchez, A. \&
  Moreno, Y.
\newblock Complex cooperative networks from evolutionary preferential
  attachment.
\newblock \emph{PLoS ONE} \textbf{3}, e2449 (2008).

\bibitem[{Poncela \emph{et~al.}(2009{\natexlab{b}})Poncela,
  G{\'o}mez-Garde{\~n}es, Traulsen \& Moreno}]{poncela_njp09}
Poncela, J., G{\'o}mez-Garde{\~n}es, J., Traulsen, A. \& Moreno, Y.
\newblock Evolutionary game dynamics in a growing structured population.
\newblock \emph{New J. Phys.} \textbf{11}, 083031 (2009{\natexlab{b}}).

\bibitem[{Szolnoki \& Perc(2009{\natexlab{a}})}]{szolnoki_epl09}
Szolnoki, A. \& Perc, M.
\newblock Resolving social dilemmas on evolving random networks.
\newblock \emph{EPL} \textbf{86}, 30007 (2009{\natexlab{a}}).

\bibitem[{Szolnoki \& Perc(2009{\natexlab{b}})}]{szolnoki_njp09}
Szolnoki, A. \& Perc, M.
\newblock Emergence of multilevel selection in the prisoner's dilemma game on
  coevolving random networks.
\newblock \emph{New J. Phys.} \textbf{11}, 093033 (2009{\natexlab{b}}).

\bibitem[{Zhang \emph{et~al.}(2011)Zhang, Zhang, Xie, Wang \&
  Perc}]{zhang_cy_pone11}
Zhang, C., Zhang, J., Xie, G., Wang, L. \& Perc, M.
\newblock Evolution of Interactions and Cooperation in the Spatial Prisoner's
  Dilemma Game.
\newblock \emph{PLoS ONE} \textbf{6}, e26724 (2011).

\bibitem[{Rand \emph{et~al.}(2011)Rand, Arbesman \& Christakis}]{rand_pnas11}
Rand, D.~G., Arbesman, S. \& Christakis, N.~A.
\newblock Dynamic social networks promote cooperation in experiments with
  humans.
\newblock \emph{Proc. Natl. Acad. Sci. USA} \textbf{108}, 19193--19198 (2011).

\bibitem[{Gracia-L{\'a}zaro \emph{et~al.}(2012{\natexlab{a}})Gracia-L{\'a}zaro,
  Cuesta, S{\'a}nchez \& Moreno}]{gracia-lazaro_srep12}
Gracia-L{\'a}zaro, C., Cuesta, J., S{\'a}nchez, A. \& Moreno, Y.
\newblock Human behavior in Prisoner's Dilemma experiments suppresses network
  reciprocity.
\newblock \emph{Sci. Rep.} \textbf{2}, 325 (2012{\natexlab{a}}).

\bibitem[{Gracia-L{\'a}zaro
  \emph{et~al.}(2012{\natexlab{b}})}]{gracia-lazaro_pnas12}
Gracia-L{\'a}zaro, C. \emph{et~al.}
\newblock Heterogeneous networks do not promote cooperation when humans play a
  prisoner's dilemma.
\newblock \emph{Proc. Natl. Acad. Sci. USA} \textbf{109}, 12922--12926
  (2012{\natexlab{b}}).

\bibitem[{Szab{\'o} \emph{et~al.}(2010)Szab{\'o}, Szolnoki, Varga \&
  Hanusovszky}]{szabo_pre10}
Szab{\'o}, G., Szolnoki, A., Varga, M. \& Hanusovszky, L.
\newblock Ordering in spatial evolutionary games for pairwise collective
  strategy updates.
\newblock \emph{Phys. Rev. E} \textbf{80}, 026110 (2010).

\bibitem[{Lorenz \emph{et~al.}(2011)Lorenz, Rauhut, Schweitzer \&
  Helbing}]{lorenz_pnas11}
Lorenz, J., Rauhut, H., Schweitzer, F. \& Helbing, D.
\newblock How social influence can undermine the wisdom of crowd effect.
\newblock \emph{Proc. Natl. Acad. Sci. USA} \textbf{108}, 9020--9025 (2011).

\bibitem[{Huberman \& Glance(1993)}]{huberman_pnas93}
Huberman, B. \& Glance, N.
\newblock Evolutionary games and computer simulations.
\newblock \emph{Proc. Natl. Acad. Sci. USA} \textbf{90}, 7716--7718 (1993).

\bibitem[{Traulsen \emph{et~al.}(2010)Traulsen, Semmann, Sommerfeld, Krambeck
  \& Milinski}]{traulsen_pnas10}
Traulsen, A., Semmann, D., Sommerfeld, R.~D., Krambeck, H.-J. \& Milinski, M.
\newblock Human strategy updating in evolutionary games.
\newblock \emph{Proc. Natl. Acad. Sci. USA} \textbf{107}, 2962--2966 (2010).

\bibitem[{Szab{\'o} \emph{et~al.}(2005)Szab{\'o}, Vukov \&
  Szolnoki}]{szabo_pre05}
Szab{\'o}, G., Vukov, J. \& Szolnoki, A.
\newblock Phase diagrams for an evolutionary prisoner's dilemma game on
  two-dimensional lattices.
\newblock \emph{Phys. Rev. E} \textbf{72}, 047107 (2005).

\bibitem[{Perc(2006{\natexlab{b}})}]{perc_njp06a}
Perc, M.
\newblock Coherence resonance in spatial prisoner's dilemma game.
\newblock \emph{New J. Phys.} \textbf{8}, 22 (2006{\natexlab{b}}).

\bibitem[{Vukov \emph{et~al.}(2006)Vukov, Szab{\'o} \& Szolnoki}]{vukov_pre06}
Vukov, J., Szab{\'o}, G. \& Szolnoki, A.
\newblock Cooperation in the noisy case: Prisoner's dilemma game on two types
  of regular random graphs.
\newblock \emph{Phys. Rev. E} \textbf{73}, 067103 (2006).

\bibitem[{Szolnoki \emph{et~al.}(2009)Szolnoki, Perc \&
  Szab{\'o}}]{szolnoki_pre09c}
Szolnoki, A., Perc, M. \& Szab{\'o}, G.
\newblock Topology-independent impact of noise on cooperation in spatial public
  goods games.
\newblock \emph{Phys. Rev. E} \textbf{80}, 056109 (2009).

\bibitem[{Hauert \& Szab{\'o}(2005)}]{hauert_ajp05}
Hauert, C. \& Szab{\'o}, G.
\newblock Game theory and physics.
\newblock \emph{Am. J. Phys.} \textbf{73}, 405--414 (2005).

\bibitem[{Szab{\'o} \emph{et~al.}(2004)Szab{\'o}, Szolnoki \&
  Izs{\'a}k}]{szabo_jpa04}
Szab{\'o}, G., Szolnoki, A. \& Izs{\'a}k, R.
\newblock Rock-scissors-paper game on regular small-world networks.
\newblock \emph{J. Phys. A: Math. Gen.} \textbf{37}, 2599--2609 (2004).

\bibitem[{Barab{\'a}si \& Albert(1999)}]{barabasi_s99}
Barab{\'a}si, A.-L. \& Albert, R.
\newblock Emergence of scaling in random networks.
\newblock \emph{Science} \textbf{286}, 509--512 (1999).

\bibitem[{Press \& Dyson(2012)}]{press_pnas12}
Press, W.~H. \& Dyson, F.~J.
\newblock Iterated Prisoner's Dilemma contains strategies that dominate any
  evolutionary opponent.
\newblock \emph{Proc. Natl. Acad. Sci. USA} \textbf{109}, 10409--10413 (2012).

\bibitem[{Stewart \& Plotkin(2012)}]{stewart_pnas12}
Stewart, A.~J. \& Plotkin, J.~B.
\newblock Extortion and cooperation in the Prisoner's Dilemma.
\newblock \emph{Proc. Natl. Acad. Sci. USA} \textbf{109}, 10134--10135 (2012).

\bibitem[{Nowak \& Sigmund(1993)}]{nowak_n93}
Nowak, M.~A. \& Sigmund, K.
\newblock A strategy of win-stay, lose-shift that outperforms tit-for-tat in
  the Prisoner's Dilemma game.
\newblock \emph{Nature} \textbf{364}, 56--58 (1993).

\bibitem[{Macy \& Flache(2002)}]{macy_pnas02}
Macy, M.~W. \& Flache, A.
\newblock Learning dynamics in social dilemmas.
\newblock \emph{Proc. Natl. Acad. Sci. USA} \textbf{99}, 7229--7236 (2002).

\bibitem[{Liu \emph{et~al.}(2011)}]{liu_yk_epl11}
Liu, Y. \emph{et~al.}
\newblock Aspiration-based learning promotes cooperation in spatial
  prisoner's dilemma games.
\newblock \emph{EPL} \textbf{94}, 60002 (2011).

\bibitem[{Liu \emph{et~al.}(2012)Liu, Chen, Zhang, Wang \&
  Perc}]{liu_yk_pone12}
Liu, Y., Chen, X., Zhang, L., Wang, L. \& Perc, M.
\newblock Win-Stay-Lose-Learn Promotes Cooperation in the Spatial Prisoner's
  Dilemma Game.
\newblock \emph{PLoS ONE} \textbf{7}, e30689 (2012).

\bibitem[{Surowiecki(2004)}]{surowiecki_04}
Surowiecki, J.
\newblock \emph{The Wisdom of Crowds: Why the Many Are Smarter than the Few and
  How Collective Wisdom Shapes Business, Economies, Societies, and Nations}
  (Random House, New York, US, 2004).

\bibitem[{Szolnoki \emph{et~al.}(2012)Szolnoki, Wang \& Perc}]{szolnoki_srep12}
Szolnoki, A., Wang, Z. \& Perc, M.
\newblock Wisdom of groups promotes cooperation in evolutionary social
  dilemmas.
\newblock \emph{Sci. Rep.} \textbf{2}, 576 (2012).

\bibitem[{Szab{\'o} \& T{\H{o}}ke(1998)}]{szabo_pre98}
Szab{\'o}, G. \& T{\H{o}}ke, C.
\newblock Evolutionary prisoner's dilemma game on a square lattice.
\newblock \emph{Phys. Rev. E} \textbf{58}, 69--73 (1998).

\bibitem[{Fu \emph{et~al.}(2010)Fu, Nowak \& Hauert}]{fu_jtb10}
Fu, F., Nowak, M.~A. \& Hauert, C.
\newblock Invasion and expansion of cooperators in lattice populations:
  Prisoner's dilemma vs. Snowdrift games.
\newblock \emph{J. Theor. Biol.} \textbf{266}, 358--366 (2010).

\end{thebibliography}
\end{document}